\documentclass[a4paper,aps,byrevtex,preprint,secnumarabic,tightenlines,nofootnoteinbib]{revtex4}
\usepackage{hyperref}
\usepackage{graphicx}
\begin{document}
\newcommand{\rzc}{R\lowercase{b}$_2$Z\lowercase{n}C\lowercase{l}$_4$}
\newcommand{\sto}{S\lowercase{r}T\lowercase{i}O$_3$}
\newcommand{\rcf}{R\lowercase{b}$_2$C\lowercase{o}F$_4$}
\newcommand{\utc}{T_c}
\newcommand{\isdd}{Ising--2D}
\newcommand{\istd}{Ising--3D}
\newcommand{\heistd}{Heisenberg--3D}
\newcommand{\tdxy}{XY--3D}
\newcommand{\ud}{\mathrm{d}}
\newcommand{\ure}{\epsilon}
\newcommand{\ukk}{\kappa}
\newcommand{\ukkk}{\kappa'}
\newcommand{\etal}{\emph{et al.}}
\newcommand{\eref}[1]{equation~(\ref{#1})}
\newcommand{\fref}[1]{figure~\ref{#1}}
\newcommand{\Fref}[1]{Figure~\ref{#1}}
\newcommand{\tref}[1]{table~\ref{#1}}
\newcommand{\sref}[1]{section~\ref{#1}}
\title[The order parameter-entropy]{The order parameter--entropy relation in some universal classes: experimental evidence}

\author{J M Mart\'{\i}n-Olalla}
\email{olalla@us.es}
\author{F J Romero}
\author{S Ramos}
\author{M C Gallardo}
\affiliation{Departamento de F\'{\i}sica de la Materia Condensada. ICMSE-Universidad de Sevilla. \\ Apartado de Correos 1065. E-41080 Sevilla. Spain}
\author{J M Perez-Mato}
\affiliation{Departamento de F\'{\i}sica de la Materia Condensada. Universidad del Pa\'{\i}s Vasco-EHU. Campus de Leioa. \\ Apartado de Correos 644 E-48080 Bilbao. Spain}
\author{E K H Salje}
\affiliation{Earth Science Department. University of Cambridge. \\ Downing Street. CB23EQ Cambridge. United Kingdom}

\bibliographystyle{apsrev}
\begin{abstract}
The asymptotic behaviour near phase transitions can be suitably characterized by the scaling of $\Delta s/Q^2$ with $\ure=1-T/T_c$, where $\Delta s$ is the excess entropy and $Q$ is the order parameter. As $\Delta s$ is obtained by integration of the experimental excess specific heat of the transition $\Delta c$, it displays little experimental noise so that the curve $\log(\Delta s/Q^2)$ versus $\log\ure$ is better constrained than, say, $\log\Delta c$ versus $\log\ure$. The behaviour of $\Delta s/Q^2$ for different universality classes is presented and compared. In all cases, it clearly deviates from being a constant. The determination of this function can then be an effective method to distinguish asymptotic critical behaviour. For comparison, experimental data for three very different systems, \rcf{}, \rzc{} and \sto{}, are analysed under this approach.  In \sto{}, the function $\Delta s/Q^2$ does not deviate within experimental resolution from a straight line so that, although Q can be fitted with a non mean-field exponent, the data can be explained by a classical Landau mean-field behaviour. In contrast, the behaviour of $\Delta s/Q^2$ for the antiferromagnetic transition in \rcf{} and the normal-incommensurate phase transition in \rzc{} is fully consistent with the asymptotic critical behaviour of the universality class corresponding to each case. This analysis supports, therefore, the claim that incommensurate phase transitions in general, and the A$_2$BX$_4$ compounds in particular, in contrast with most structural phase transitions, have critical regions large enough to be observable.
\end{abstract}

\maketitle

\section{Introduction}
\label{sec:intro}
                The observation of fluctuation-driven critical phenomena in structural phase transitions is subject of long and persistent controversies. Although power law behaviours near phase transitions can be ``universal'', the temperature range where this asymptotic behaviour can be observed within experimental resolution (the so-called asymptotic or critical region) is not. It depends in general on the particular features of each system and can even vary for different physical quantities. It is no wonder, therefore, that the size of the critical region in structural phase transitions has been a matter of discussion for several decades. In general, one should expect it to be smaller than in the pure order-disorder systems, the spin hamiltonians, that represent the corresponding universality classes. However, the Ginzburg criterion has been equally invoked to predict typical critical regions of size too small to be of any experimental significance,\cite{ginzburg-ferro-87} as to justify critical regions of the order of 0.1 in reduced temperature ($T/T_c$)\cite{muller-prl-71} (an interval that would greatly surpass that of the actual spin-hamiltonian representatives of some of the universality classes). In addition, structural defects may produce strong deviations from the universal asymptotic  laws predicted for ideally perfect crystals.\cite{levanyuk-spj-79,scott-ferro-81,marro-prb-86}

Authors expecting insignificantly small critical regions have argued repeatedly that the experimental observation of power laws around the phase transition with non-mean-field exponents may have a rather simple explanation, e.g. in terms of the Landau theory when sixth-order or higher order terms are taken into account in the Landau expansion\cite{ginzburg-ferro-87}. For instance, a 2-4-6 Landau potential:

\begin{equation}
  \label{eq:landaupot}
  \Delta G=\frac{1}{2}A \ure Q^2+\frac{1}{4}BQ^4+\frac{1}{6}CQ^6
\end{equation}                                                          
with temperature independent parameters $A,B,C$ and $\ure=1-T/T_c$ yields indeed a mean field critical behaviour for the order parameter ($Q\propto\ure^{1/2}$) with a mean-field exponent $\beta=1/2$, but the asymptotic behaviour is limited to $\ure\ll B^2/4AC$. This temperature interval can be beyond any observation if the fourth order coefficient is small enough, a typical situation happening if the system is close to a tricritical point. In $\ure$-intervals of the order of 0.01-0.1, or even larger, the behaviour of the order parameter can then be fitted to power laws with ``effective'' exponents which are intermediate between the mean field value 1/2 and the one corresponding to mean field tricritical behaviour (1/4). 

        More generally, it became clear that any experimental observation of expected asymptotic critical behaviour for the order parameter requires further confirmation through cross-examination of its consistency with the behaviour of other physical parameters near the transition point. The correlation of the order parameter with the transition excess entropy has been recently used for this purpose\cite{ekhard-jpc-98}. The excess entropy $\Delta s$ is expected to be proportional to the square of the order parameter $Q$ not only in the mean field asymptotic regime ($\beta=1/2, \alpha=0$), but also for any behaviour described by a Landau potential like~\eref{eq:landaupot}. The relation would be true as long as the terms higher than quadratic in the Landau potential are temperature independent and the quadratic term has a linear temperature dependence. This particular relation between entropy and order parameter is in principle broken in the asymptotic regime for all other universality classes. Its experimental check can be then an effective way to distinguish between genuine fluctuation-driven critical behaviour and pure Landau behaviour. Indeed, this proportionality relation has recently been shown for \sto, in contradiction with a common belief that this compound exhibits in a significant temperature interval asymptotic critical behaviour corresponding to the \heistd\ or the \istd\ universality class\cite{ekhard-jpc-98,chrosch-jpc-98,stuart-pt-99}. The analysis of $\Delta s$, instead of checking directly the expected critical behaviour of the excess specific heat $\Delta c$, has the fundamental advantage of using an integrated quantity with a smooth temperature dependence that is independent of experimental statistical errors in the measurement of $\Delta c$.
 
In the present paper, the approach of comparing $\Delta s$ and $Q^2$ for analysing the thermal behaviour at phase transitions is further investigated and applied to several materials. The expected behaviour of $\Delta s/Q^2$ for the different universality classes and the Landau case, is discussed and compared with that derived from experimental data for \rcf{} (antiferromagnetic phase transition), \rzc{} (normal-incommensurate phase transition) and \sto{} (antiferrodistortive transition).

\section{ Entropy-order parameter relation in the critical region}
\label{sec:entropy-order}

The asymptotic entropy-order parameter relation for the different universality classes follows from the critical laws for specific heat and order parameter. It can be expressed as 
\begin{equation}
  \label{eq:index}
\frac{\Delta s}{Q^2}\propto \ure^\ukk \quad\quad  \Delta s\propto Q^{2\ukkk} 
\end{equation}
with 
\begin{equation}
  \label{eq:index2}
  \ukk=
\left\{\begin{array}{ll}
1-\alpha-2\beta&\alpha>0\\
1-2\beta&\alpha\leq0\\
  \end{array}\right.
\end{equation}
and
\begin{equation}
  \label{eq:index3}
  \ukkk=
\left\{\begin{array}{ll}
\frac{1-\alpha}{2\beta}&\alpha>0\\
\frac{1}{2\beta}&\alpha\leq0\\
  \end{array}\right.
\end{equation}
          
In order to get these expressions the normal asymptotic behaviour of the order parameter, characterized by the exponent $\beta$, is assumed while that of the excess specific heat is supposed to be:
    \begin{equation}
  \label{eq:assum}
  \Delta c\propto\frac{\ure^{-\alpha}-1}{\alpha}
\end{equation}

Note that the specific heat is not reduced to a pure power law as it includes the cases of negative $\alpha$ values and the logarithmic divergence for $\alpha\to0$.\cite{heller-rpp-1967,griffiths-pr-1967}
\begin{table*}[tbp]
  \centering
  \begin{tabular}{|l|ll|llll|ll|}
\hline
Class&$n$&$d$&$\alpha$&$\beta$&$\kappa$&$\kappa'$&$\ukk_{e_T}$&$\ukkk_{e_T}$\\
\hline
Landau (2-4)&-&-&\phantom{ }0&0.5&0&1&0&1\\
Landau (2-6)&-&-&\phantom{ }0.5&0.25&0&1&0&1\\
\isdd&1&2&\phantom{ }0&0.125&0.75&4&0.598&3.39\\
\istd&1&3&\phantom{ }0.12&0.31&0.26&1.42&0.163&1.26\\
\tdxy&2&3&-0.007&0.345&0.31&1.45&0.164&1.24\\
\heistd&3&3&-0.14&0.37&0.26&1.31&0.182&1.25\\
\hline
  \end{tabular}
   \caption{The critical exponents for several universal classes. The dimensionality of the ``interaction'' is labeled $d$. The dimensionality of order parameter is labeled $n$. Exponent $\alpha$ is that of the specific heat, $\beta$ is  the order parameter exponent, In the limit $\ure\to0$, $\kappa$ and $\ukkk$ are given by \eref{eq:index2} and \eref{eq:index3}. $\ukk_{e_T}$ stands for the  ``effective'' exponent obtained in an $\ure$-interval of $10^{-3}-10^{-2}$ for the theoretical assymptotic behaviour of the models.}
  \label{tab:modelos}
\end{table*}

\begin{figure}[t]
  \centering
  \includegraphics[angle=270,width=0.7\columnwidth]{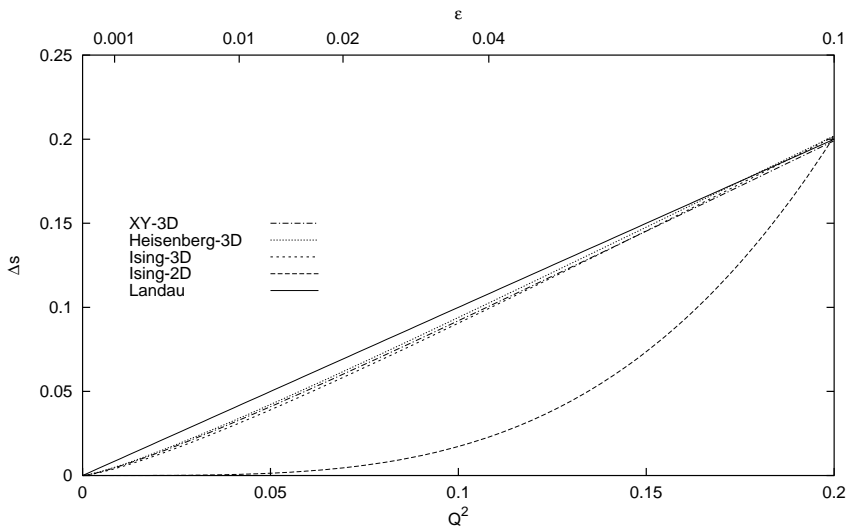}
 \caption{Theoretical behaviour of the excess entropy vs the square of the order parameter for various universality classes. Landau mean field theory gives a straight line passing through origin. Order parameter is normalized to unity at $\ure=1$, while entropies have been renormalized so as to match at $Q^2=0.2$.  Upper labels stand for $\ure$ in the \tdxy\ universality class, they are also approximately valid for \heistd\ and \istd\ but not for \isdd\ or for Landau behaviour.}
  \label{fig:teordsqd}
\end{figure}

The resulting exponents $\ukk$ and $\ukkk$ for the different universality classes and the Landau behaviour are listed in \tref{tab:modelos}. Non-classical exponents $\ukk$ and $\ukkk$ differ significantly from the mean-field/Landau trivial values associated with the above mentioned proportionality law. Nevertheless, these deviations may be rather difficult to detect in a typical $\Delta s$ vs. $Q^2$ plot. In \fref{fig:teordsqd}, we present these plots for the different universality classes assuming that, in an $\ure$ interval of 0.1 and within experimental resolution, both the order parameter and the excess specific heat follow their asymptotic critical behaviour (see \eref{eq:assum} for that of the excess specific heat).

The excess entropy in \fref{fig:teordsqd} has been calculated by numerical integration of the expression:
 
\begin{equation}
  \label{eq:entro}
   \Delta s=\int\limits_{T_c}^T\frac{\Delta c}{T'}\mathrm{d}T'=\int\limits_0^\ure\frac{\Delta c}{1-\ure'}\mathrm{d}\ure'
\end{equation}
this integration leads generally to smooth curves $\Delta s(T)$ even if $\Delta c$ contains significant statistical noise.

\begin{figure}[t]
  \centering
  \includegraphics[width=0.7\columnwidth]{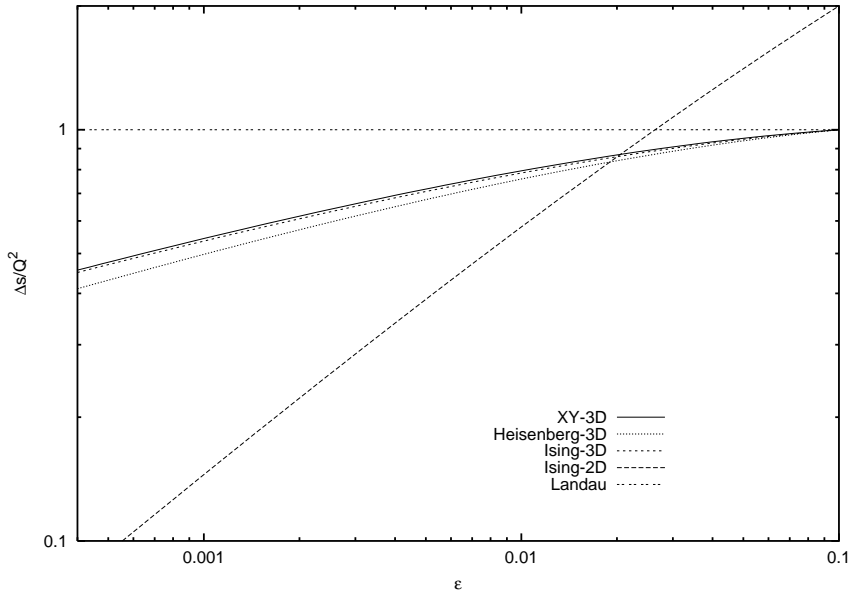}

 \caption{$\Delta s/Q^2$ versus $\ure$ in a log-log plot for the same data used for curves of \fref{fig:teordsqd}. The magnitude represented in $y$-axis is normalized to unity at $\ure=0.1$ except that of \isdd\ which has been arbitrarily shifted upwards for the sake of clarity.}

  \label{fig:teordsqdb}
\end{figure}

Classical Landau behaviour is represented in \fref{fig:teordsqd} by a straight line through the origin. The \istd , \heistd\ and \tdxy\ classes yield curves different but close to a straight line. Therefore, in practice, the deviation from the Landau behaviour is difficult to assess quantitatively through these plots. Only in the case of the \isdd\ class the curve, being a fourth order parabola, clearly deviates from the proportionality law. The different relationships between $\Delta s$ and $Q^2$ become more obvious in a log-log plot of $\Delta s/Q^2$ vs $\ure$. The plots corresponding to the curves of \fref{fig:teordsqd} are presented in \fref{fig:teordsqdb}. The deviation from the horizontal Landau behaviour is then obvious for all universality classes. The larger deviation is seen for the \isdd\ case, while all other universality classes lead again to more similar curves. It also becomes evident in the figure that although the curves have been obtained assuming perfect asymptotic behaviour for $\Delta c$ and $Q$ up to $\ure=0.1$, the magnitude $\Delta s/Q^2$ is far from being in the asymptotic regime represented by the power law of \eref{eq:index2} even at $\ure$ values so low as 0.0005. The curves deviate rather smoothly from straight lines and in fact their approximate slopes are far from the asymptotic values given by \eref{eq:index2} and listed in \tref{tab:modelos}. Indeed, the ``critical'' region for the excess entropy is much smaller than for $\Delta c$, since the integration represented by \eref{eq:entro} yields the sum of several competing terms with similar but different powers of $\ure$. Hence, according to \fref{fig:teordsqdb}, the asymptotic behaviour for $\Delta s/Q^2$ represented by \eref{eq:index} is not experimentally accessible in any case. Nevertheless, the curves in \fref{fig:teordsqdb} can be considered ``universal'' in the sense that they should be satisfied for any system so long as its order parameter and specific heat are following the asymptotic critical behaviour of the corresponding universality class. In the $\ure$ interval [0.001-0.01] the curves can be approximated to straight lines, their slopes correspond then to ``universal'' effective exponents $\ukk_e$ for the approximate law $\Delta s/Q^2\propto\ure^{\ukk_{e_T}}$ restricted to this  temperature interval. The exponents $\ukk_{e_T}$ obtained from the linear fit of the curves in \fref{fig:teordsqdb} in the mentioned temperature interval are listed in \tref{tab:modelos}.

\section{Experimental cases}
\label{sec:experimental}

We consider now under this approach three very different systems: strontium titanate (\sto), rubidium tetrafluorocobaltate (\rcf)  and rubidium tetrachlorozincate (\rzc). 

\subsection{\sto}
\label{sec:sto}
The antiferrodistortive phase transition of \sto{} has been widely considered one of the few examples of a structural phase transition with a clear observable critical region. M\"uller \etal\cite{muller-prl-71} could fit their EPR data assuming a power law for the order parameter with $\beta=0.35$ in an $\ure$ interval of the order of 0.1.  This behaviour was then identified as the expected asymptotic behaviour of the \heistd\ class. Recently,  the excess entropy was determined and compared with the order parameter data; a mean field linear correlation was observed\cite{ekhard-jpc-98}, thus the critical region might be beyond observation. For comparison with earlier specific heat data see Ref.~\cite{carmen-jpc-02}. 

For the sake of comparison with the two following systems, the $\Delta s$-$Q^2$ relation obtained in Ref.~\cite{ekhard-jpc-98} for \sto{} is reproduced in \fref{fig:sto} in a format similar to next figures.
\begin{figure}
  \centering
  \includegraphics[width=0.7\columnwidth]{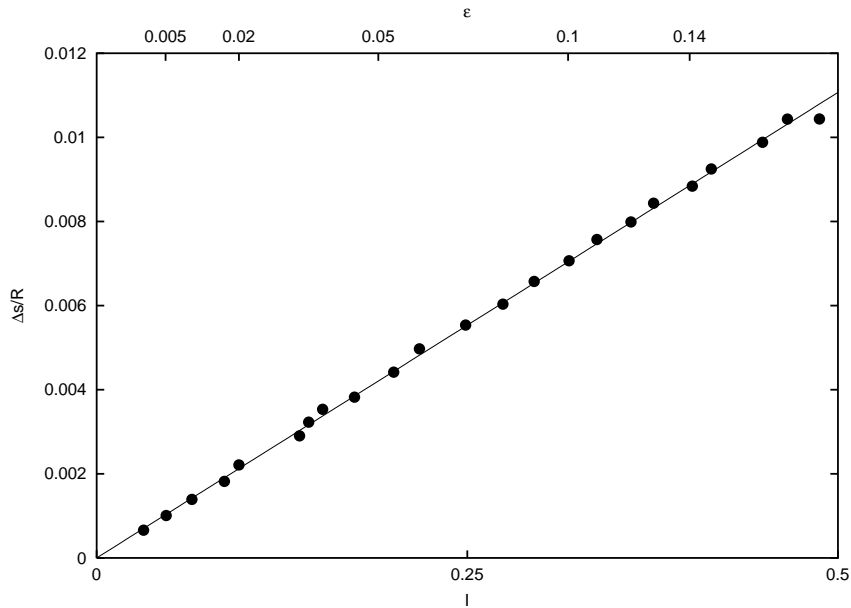}
  \caption{Excess entropy versus excess intensity of X-ray difraction experiments around the \sto\ phase transition at about $106\,\mathrm{K}$. In the upper axis approximate values for reduced temperature are indicated}
  \label{fig:sto}
\end{figure}
\subsection{\rcf}
\label{sec:rcf}
\begin{figure}[tbp]
  \centering
  \includegraphics[width=0.7\columnwidth]{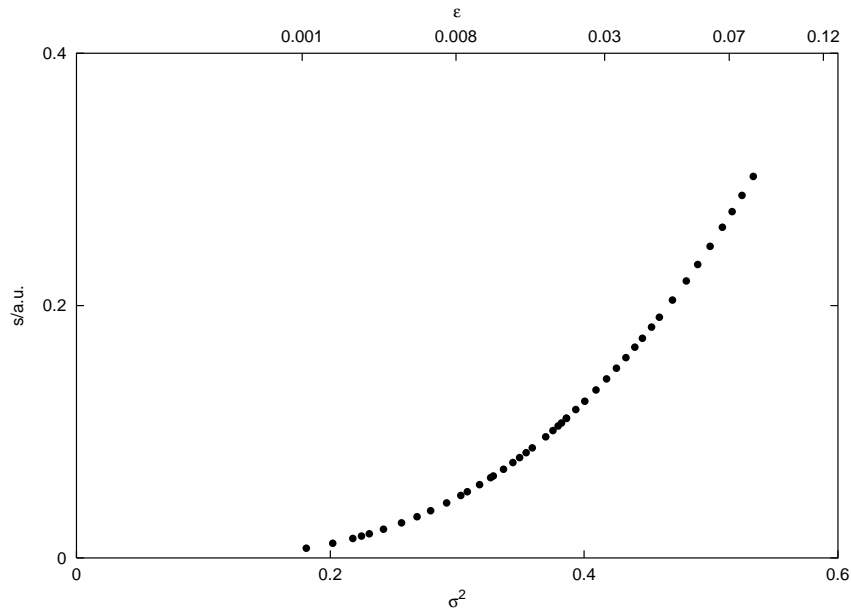}
  \caption{The excess entropy of the antiferromagnetic phase transition in \rcf\ versus the square of the reduced magnetization, as obtained using data from Samuelsen\protect\cite{samuelsen-prl-73} and Ikeda \etal\protect\cite{ikeda-jpsj-76}  Upper labels represent reduced temperature.}
  \label{fig:rcfa}
\end{figure}

This compound exhibits an antiferromagnetic phase transition at $107\,\mathrm{K}$. This system is believed to belong to the Ising-2d universality class. The structure of \rcf\ is isomorphic to that of K$_2$NiF$_4$ which is a well known two-dimen\-sional anti-ferromag\-netic system. The two-dimen\-sional character is related to the presence of coupled anti-ferromag\-netic planes separated by planes with weak interplanar effective interactions. Specific heat \cite{ikeda-jpsj-76} and the reduced sublattice magnetization (order parameter) obtained from neutron scattering experiments\cite{samuelsen-prl-73} have been reported to be consistent with the critical behaviour predicted for this universality class. In \fref{fig:rcfa} we present the correlation of the excess entropy entropy $\Delta s$ and the square of the order parameter (the reduced sublattice magnetization $\sigma$) for this material. The entropy has been calculated upon integration of the specific heat measured by Ikeda \etal\cite{ikeda-jpsj-76}. The magnetization was determined from neutron scattering experiments by Samuelsen\cite{samuelsen-prl-73}. The figure shows clearly how the entropy fails to be proportional to the square of the order parameter.

\subsection{\rzc}
\label{sec:rzc}
\rzc\ presents an orthorhombic (Pnma) phase\cite{cummins-pr-90} abo\-ve $T_{c}\sim 305\,\mathrm{K}$. At this temperature it transforms through a continuous phase transition into an incommensurately modulated crystal with a modulation wave vector close to one third of  $\mathbf{a}^{*}.$ Further below, at $195\,\mathrm{K}$ the modulation wave vector locks into the exact value $1/3\:\mathbf{a}^*$, and the incommensurate phase transforms into an orthorrhombic commensurate ferroelectric phase with a three-fold unit cell along the modulation. This compound belongs to the family of type A$_2$BX$_4$, where this incommensurate instability is rather common, and has been extensively studied in order to test the theoretical predictions on incommensurate phase transitions. In 1978 Bruce and Cowley\cite{bruce-jpc-78} argued that this transition should belong to the \tdxy\ universality class. Since then, many workers have related their results to what is expected in the critical zone for this universality class. For instance, Walisch \etal\cite{walish-prb-89} accurately determined from quadrupole perturbed NMR experiments the critical exponent of the order parameter with full agreement with the theoretical value for the \tdxy\ class. According to these results the critical zone would reach up to $\ure$ values of the order of 0.07. Diffraction synchrotron measurements\cite{zinkin-prb-96} confirmed these results and a critical region of the order of 0.1.  Specific heat was also reported\cite{chen-prb-90} to agree expected critical behaviour in a range up to $\ure\sim0.2$. More recently, Haga \etal\cite{haga-jpsj-95} made a more comprehensive study of the specific heat and fitted their data to an expression which includes corrections up to second order of the asymptotic law. The fitted $\ure$ interval reached in this case up to $\ure=0.1$. The existence of observable large critical regions in incommensurate phase transitions, in comparison with normal structural phase transitions, was predicted and justified.\cite{patashinskii-spj-74,prokovskii-aip-79} However, other authors maintain that there is no theoretical ground for expecting in incommensurate systems larger critical regions than in other structural phase transitions\cite{ginzburg-ferro-87,sandler-spss-83} and have interpreted for instance other NMR experiments in this compounds  within a Landau approach\cite{fajdiga-prl-92}.
 
        For the present study, a new independent measurement of the specific heat of \rzc\ was performed in a conduction calorimeter. The calorimeter has been previously described \cite{jaime-jsi-87} and has been used successfully for the measurements of thermal properties of single-crystals\cite{jose-pt-97}. A single crystal of  $1.078\,\mathrm{g}$  ($2.850\,\mathrm{mmol}$) was measured. The sample was grown from aqueous solution and was annealed at the high symmetry phase and then cooled down at a low rate of $0.4\mathrm{Kh^{-1}}$ while the heat capacity of the sample was recorded. The heat capacity agrees previously reported data\cite{haga-jpsj-95,echarri-ferro-80,atake-chemther-83} and they show a $\lambda$-shaped peak at about $304\,\mathrm{K}$.

\begin{figure}[tbp]
  \centering
  \includegraphics[width=0.7\columnwidth]{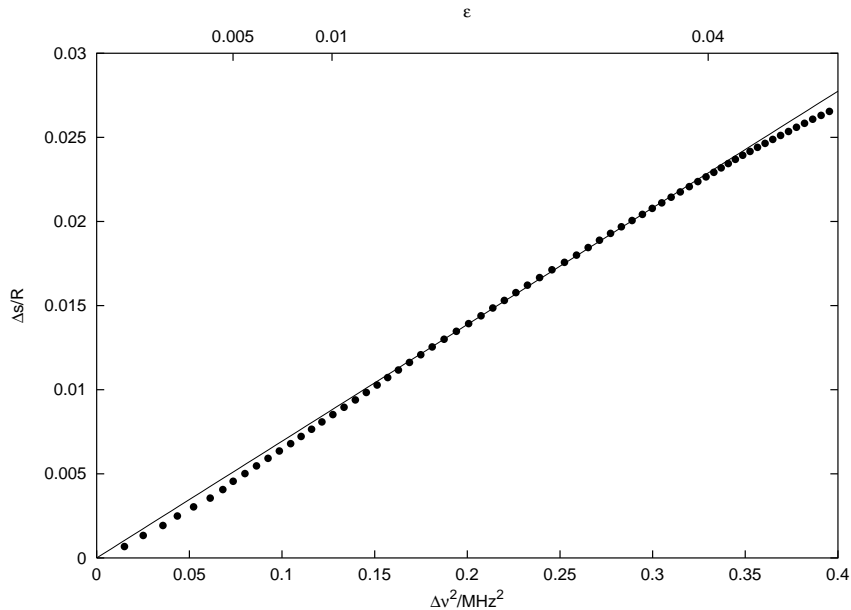}
  \caption{Excess entropy versus square of the NMR frequency distribution width $\Delta\nu$ proportional to the order parameter in \rzc. The straight line represents the Landau mean field behaviour. Labels at upper $x$-axis represents $\ure$.}
  \label{fig:rzca}
\end{figure}

 The background contribution to the specific heat was determined by fitting a smooth polynomial function to data above $\utc$ and data well below the lock-in phase transition located at $T_i=195\,\mathrm{K}$. By subtracting the background contribution, the excess specific heat was determined and the excess entropy versus temperature was obtained through integration. This latter could then be compared with the temperature evolution of the width $\Delta\nu$ of the NMR frequency distribution measured by Walisch \etal\cite{walish-prb-89}, which is a quantity proportional to the order parameter. Due to the sharpness of the specific heat anomaly we have assumed that $T_c$ corresponds to the temperature of maximum specific heat. The temperature origin of the NMR data was slightly modified to make $T_c$ coincide. Excess entropy was set to zero at $\ure=0$. Finally, an interpolation function was used to get the values of the NMR frequency distribution width $\Delta\nu$ at the temperatures values were entropy was determined. \Fref{fig:rzca} presents the correlation between both quantities in a $\Delta s$ versus $\Delta\nu^2$ graph. The deviation from a mean field linear behaviour is rather small, but as shown in \fref{fig:teordsqd}, \tdxy\ asymptotic behaviour implies in this representation small changes with respect to the Landau mean field behaviour. 

\section{Discussion}
\label{sec:discu}

\subsection{Comparative analysis of experimental data}
\label{sec:comp}
\begin{figure}[tbp]
  \centering
  \includegraphics[width=0.7\columnwidth]{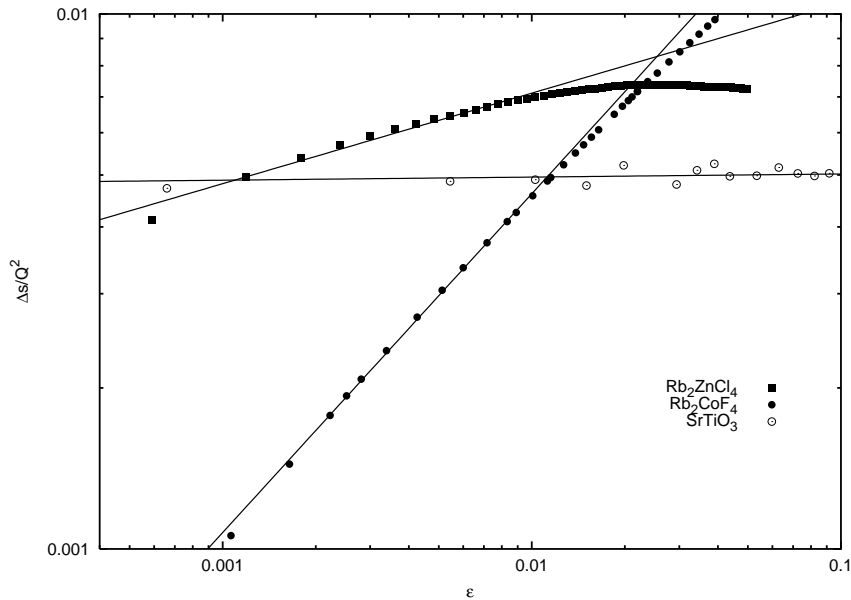}
  \caption{Log-log plot of the ratio $\Delta s/Q^2$ ---in arbitrary units--- as a function of $\ure$ for \rzc , \sto , and \rcf . Straight lines show the linear fitting done in the intervals indicated in Table 2. Resulting effective exponents $\kappa_e$ are summarized in \tref{tab:expon}.}
  \label{fig:rzcb}
\end{figure}

\Fref{fig:rzcb} depicts the log-log plot of $\Delta s/Q^2$ versus $\ure$ for the three compounds discussed in \sref{sec:experimental}. One can clearly see the deviation of both \rzc{} and \rcf{} from the Landau mean field behaviour, in contrast with the case of \sto{}. In accordance with the simulations in \sref{sec:entropy-order}, the curves in general can be approximated to straight lines when reduced to the interval of smaller $\ure$ values. Their slopes yield ``experimental effective exponents'' $\ukk_e$, listed in \tref{tab:expon}, that can be compared with the theoretical ones listed in \tref{tab:modelos}. It can be stated that in the case of \rzc{} the experimental effective exponent $\ukk_e$ value coincides with the theoretical value of the \tdxy\ class considering its standard deviation. It should be noted  that the values of $\ukk_{e_T}$ are similar for \tdxy\ and \istd\ universality classes. However, the value $\beta$ that fits the order parameter behaviour is consistent with only the \tdxy\ class. The $\ukk_e$ value for \rcf{} is close to $\ukk_{e_T}$ expected for \isdd\ class, with a deviation of the order of 5\%; while in \sto{}  not only the experimental $\ukk_e$ clearly disagrees with the theoretical value  of the \heistd\ class, but its zero value indicates a classical Landau behaviour. It should be stressed that a mere deviation from the linearity between $\Delta s$ and $Q^2$ should not be taken as a proof of fluctuation-driven critical behaviour, since a generalized Landau potential with temperature dependent quartic and higher order terms could explain such behaviour. It is the quantitative agreement of this deviation (measured by the effective exponent $\ukk_e$) with that expected for the corresponding universality class ($\ukk_{e_T}$), together with the agreement of the fitted exponent $\beta$ that proofs the consistency of an interpretation in terms of asymptotic critical behaviour for both the order parameter and the specific heat.

\begin{table}[tbp]
  \centering
  \begin{tabular}{lrcccr}
\textbf{Crystal}&\textbf{$\ure$ fitting range}&\textbf{Points}&$\ukk_e$\\
\hline
\sto&$10^{-3}$--$10^{-1}$&22&$0.006(5)$\\
\rcf&$10^{-3}$--$10^{-2}$&12&$0.633(6)$\\
\rzc&$10^{-3}$--$10^{-2}$&16&$0.169(6)$\\
  \end{tabular}
  \caption{Experimental effective exponent $\ukk_e$, deduced from the log-log fit in \fref{fig:rzcb}, for the systems under study. Values of experimental $\ukk_e$ should be compared to those expected theoretically for the universality classes $\ukk_{e_T}$ listed in \tref{tab:modelos}}
  \label{tab:expon}
\end{table}

\subsection{Crossover in  \rzc}
\label{sec:rzcb}
\begin{figure}[tbp]
  \centering
  \includegraphics[width=0.7\columnwidth]{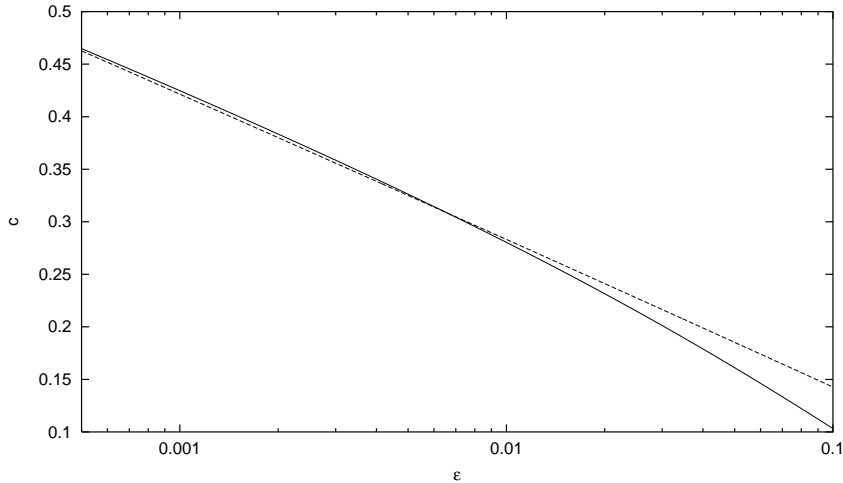}
  \caption{The solid line represents the curve (6) used by Haga \etal\protect\cite{haga-jpsj-95}to fit the excess specific heat of \rzc . The dashed line is a simple critical law as given by \eref{eq:assum} with $\alpha=-0.007$ in accordance with the \tdxy\ class.}
  \label{fig:hagakill}
\end{figure}

The curve of \rzc{} in \fref{fig:rzcb} strongly deviates from its low $\ure$ linear behaviour for $\ure>0.01$ and changes then rapidly into a near horizontal straight line. This suggests a crossover around this temperature from the \tdxy\ critical behaviour to a mean-field/Landau regime. However, this interpretation would be somehow in contradiction with the fact that the observed order parameter behaviour, used in fact for the calculation of this curve, does not suffer any kind of crossover at $\ure\sim 0.01$ and follows the $\beta$ power law with excellent agreement up to much larger $\ure$ values, of the order of 0.1.  In any case, the curve clearly deviates for $\ure>0.1$ from the theoretical curve for the \tdxy\ class in \fref{fig:teordsqdb} and this can then only be caused by the deviation beyond this $\ure$ value of the experimental $\Delta c$ from the theoretical asymptotic behaviour given by \eref{eq:assum} and used to produce \fref{fig:teordsqdb}. The reason for such deviation could be intrinsic, i.e. the critical region for the specific heat would be then of the order 0.01, while for the order parameter is about 0.1 in reduced temperature, but it is also possible that the critical region were sample dependent. In order to get a more comprehensive analysis of this phase transition we have also analyzed the measurement done by Haga \etal\cite{haga-jpsj-95} where the excess specific heat was fitted over $30\,\mathrm{K}$, i.e. 0.1 in reduced temperature, using a complex expression involving corrections up to second order of the pure asymptotic regime:
                                                                               
\begin{equation}
  \label{eq:hagafit}
  \Delta c=A\ure^{-\alpha}\left(1+D_1\ure^{\Delta_1}+D_2\ure\right)+B_c
\end{equation}
where $\alpha=-0.007$, $A=8.153, D_1=0.040, D_2=-0.026$, $B_c=8.201,\Delta_1=0.524$; where $A, D_1, D_2, B_c$ are expresed in arbitrary units. This curve is compared in \fref{fig:hagakill} with that obtained from a mere critical law as given in \eref{eq:assum}, which essentially corresponds to the neglect of the coefficients $D_1$ and $D_2$ in \eref{eq:hagafit}.

\begin{figure}[tbp]
  \centering
  \includegraphics[width=0.7\columnwidth]{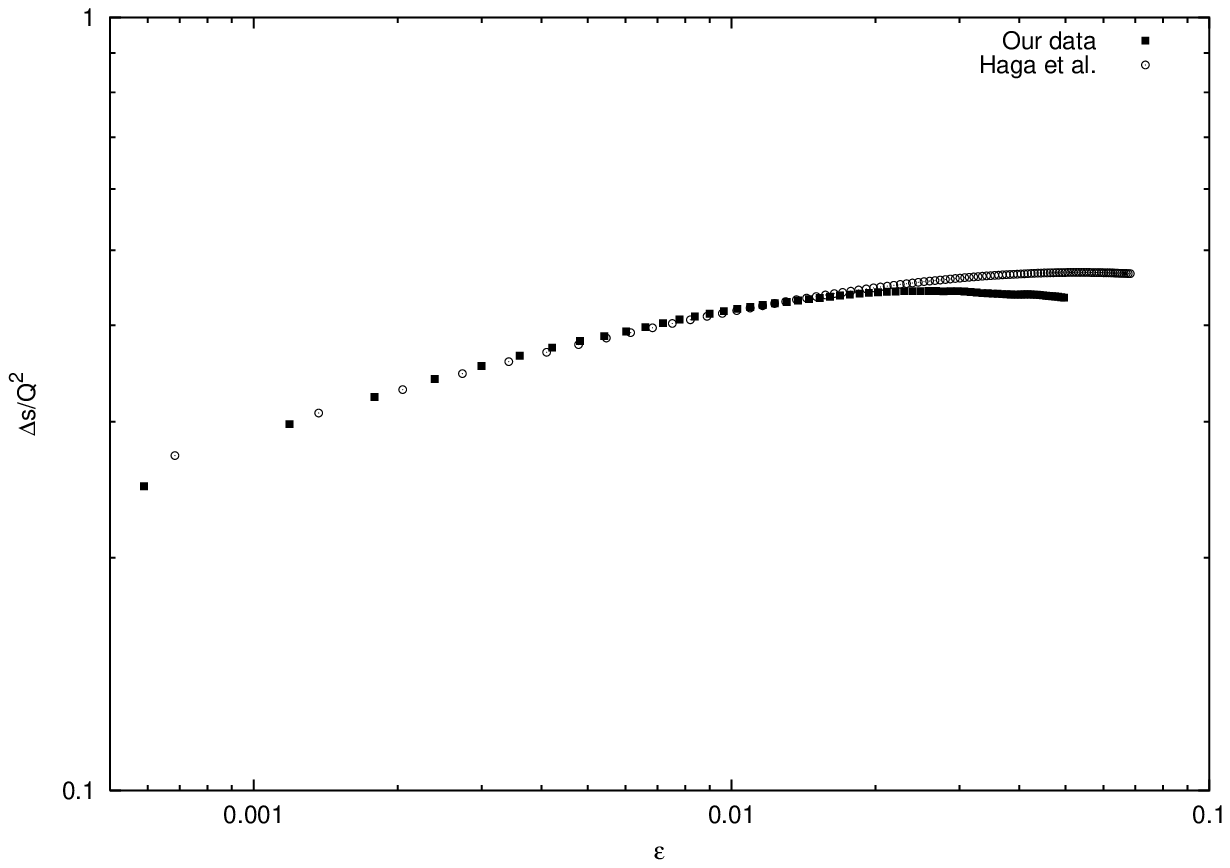}
  \caption{Log-log plot of the ratio $\Delta s/Q^2$ ---in arbitrary units--- for \rzc . We present the results for our specific heat data compared with those obtained using instead data of Haga \etal\protect\cite{haga-jpsj-95} as given by \eref{eq:hagafit}.}
  \label{fig:hagaour}
\end{figure}

The difference between both curves become obvious beyond $\ure=0.01$, showing that for lower temperatures corrections to the asymptotic behaviour are significant. For comparison, we have computed again the curve of $\Delta s/Q^2$ using as specific heat data the values given by the function \eref{eq:hagafit} as proposed by Haga. The results are presented in \fref{fig:hagaour}.

\begin{figure}[tb]
  \centering
  \includegraphics[angle=270,width=0.7\columnwidth]{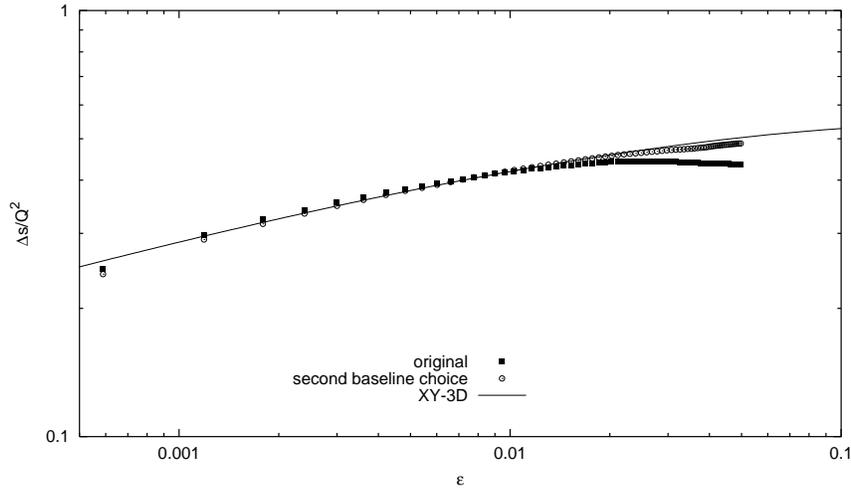}
  \caption{Effect of changing the baseline of the specific heat of \rzc{}. The solid line represents the theoretical behaviour for the asymptotic critical regime of the \tdxy\ class.}
  \label{fig:shifting}
\end{figure}

The agreement between both calculations is excellent. The approximate linear behaviour for $\ure$ values smaller than 0.01 is reproduced and its slope gives a value of around 0.15 similar to  that obtained from our specific heat analysis. On the other hand, a small shift along the $\ure$-axis of the crossover region can be observed and the behaviour for larger $\ure$ values slightly differs. These differences are most probably due to the choice of the baseline for obtaining the ``excess'' specific heat. The influence of the baseline is slight at small values of $\ure$ but becomes very important when $\ure$ increases. The results presented for \rzc{} in \fref{fig:rzcb} correspond to a baseline described by a polynomial of minimal order fitting the specific data well above and below the phase transition. This polynomial can be significantly changed, by shifting the low-temperature fitting region to higher o lower temperatures, while keeping a physically consistent baseline. The effect of this change of baseline is shown in \fref{fig:shifting}. It is slight for low $\ure$ values, where the approximate linear behaviour remains unchanged, while the crossover to what one might consider a Landau regime (the horizontal part of the curve in \fref{fig:rzcb}) becomes less pronounced.  As shown in the figure, the change is such that the calculated curve coincides with the theoretical one of the XY-3D class in a much larger $\ure$ interval. Hence, a large uncertainty exists with respect to the actual $\ure$-interval where a crossover to the Landau behaviour does take place. But, the behaviour of the $\Delta s/Q^2(\ure)$ relationship for $\ure<0.01$ confirming a XY-3D critical regime is robust with respect to modifications of the baseline.

\acknowledgements
  This work was supported by Spanish \emph{Ministerio de Ciencia y Tecnolog\'\i a} project number PB98-115.


\end{document}